%% file: main.tex
\documentclass[Afour,sageapa,times]{sagej}

\usepackage{moreverb,url}
\usepackage[colorlinks,bookmarksopen,bookmarksnumbered,citecolor=red,urlcolor=red]{hyperref}

\usepackage{multirow}
\usepackage{tabularx}
\usepackage{adjustbox}
\usepackage{xcolor,colortbl}
\usepackage{outlines}
\usepackage{todonotes}
\usepackage{subcaption}
\usepackage{tikz}
\usepackage{amsmath}
\usepackage{lineno}

\usepackage{subfiles} 

\usetikzlibrary{positioning, shapes.geometric, arrows}

\newcommand\BibTeX{{\rmfamily B\kern-.05em \textsc{i\kern-.025em b}\kern-.08em
T\kern-.1667em\lower.7ex\hbox{E}\kern-.125emX}}

\def\volumeyear{2024}

\input{generated_figures/demographics.tex}
\input{generated_figures/model.tex}

\begin{document}

\runninghead{Dekarske et al.}

\title{Dynamic Human Trust Modeling of Autonomous Agents With Varying Capability and Strategy}

\author{Jason Dekarske\affilnum{1}, Zhaodan Kong\affilnum{1}, and Sanjay Joshi\affilnum{1}}
\affiliation{\affilnum{1}University of California, Davis, Davis, CA, USA}

\email{jdekarske@ucdavis.edu}

\begin{abstract}
\subfile{sections/0abstract.tex}
\end{abstract}


\maketitle

%

\section{Introduction}\label{introduction}
\subfile{sections/1introduction.tex}

\section{Background}\label{background}
\subfile{sections/2background.tex}

\section{Experiment}\label{experiment}
\subfile{sections/3experiment.tex}

\section{Trust Model}\label{model}
\subfile{sections/4trustmodel.tex}

\section{Results}\label{results}
\subfile{sections/5results.tex}

\section{Discussion}\label{discussion}
\subfile{sections/6discussion.tex}

\section{Conclusion}\label{conclusion}
\subfile{sections/7conclusion.tex}

\begin{acks}
This material is based upon work supported by NASA (award number 80NSSC19K1052). Any opinions, findings, and conclusions or recommendations expressed in this material are those of the author(s) and do not necessarily reflect the views of the National Aeronautics and Space Administration (NASA). Thank you to the subjects which participated in this experiment. The authors appreciate helpful feedback from the members of the HOME STRI regarding trust and its applications.
\end{acks}

\section{Key Points}
\begin{itemize}
    \item Perceived trust, as measured with self-report over time, is a useful metric in cases where measuring behavioral trust is inconvenient.
    \item A cooperative visual search task with an autonomous agent elucidated changes in human trust due to agent capability, strategy, and the ordering of which these characteristics were applied.
    \item An ARIMAX time-series model outperformed linear regression. Cross-validation of ARIMAX model between groups of different characteristic ordering also showed improvement over linear regression.
\end{itemize}



\bibliographystyle{apalike}
\bibliography{grid-search.bib}

\begin{biogs}
Jason Dekarske
\begin{itemize}
    \item Current Affiliation: University of California, Davis
    \item Highest Degree Obtained: BSc Biomedical Engineering, University of Wisconsin - Madison, 2014
\end{itemize}

Zhaodan Kong
\begin{itemize}
    \item Current Affiliation: University of California, Davis
    \item Highest Degree Obtained:  PhD Aerospace Engineering and Mechanics, University of Minnesota, Twin Cities, 2012
\end{itemize}

Sanjay Joshi
\begin{itemize}
    \item Current Affiliation: University of California, Davis
    \item Highest Degree Obtained: PhD Electrical Engineering, University of California, Los Angeles, 1996
\end{itemize}
\end{biogs}


\end{document}

%% file: generated_figures/demographics.tex
\newcommand{\includedSubjects}{80}
\newcommand{\male}{22}
\newcommand{\female}{57}
\newcommand{\nonbinary}{1}

\newcommand{\games}{45}
\newcommand{\amountGames}{1 and 5 hours per week}
\newcommand{\avgAge}{20.0}
\newcommand{\stdAge}{1.5}

%% file: generated_figures/model.tex
\newcommand{\ARIMAXRMSEGone}{0.048}
\newcommand{\ARIMAXRMSEGCROSSzero}{0.094}
\newcommand{\ARIMAXRMSEGzero}{0.067}
\newcommand{\ARIMAXRMSEGCROSSone}{0.067}
\newcommand{\OLSRMSEGzero}{0.097}
\newcommand{\OLSRMSEGone}{0.062}

%% file: sections/0abstract.tex
\subsubsection{Objective}
We model the dynamic trust of human subjects in a human-autonomy-teaming screen-based task.

\subsubsection{Background}
Trust is an emerging area of study in human-robot collaboration. Many studies have looked at the issue of robot performance as a sole predictor of human trust, but this could underestimate the complexity of the interaction.

\subsubsection{Method}
Subjects were paired with autonomous agents to search an on-screen grid to determine the number of outlier objects. In each trial, a different autonomous agent with a preassigned \textit{capability} used one of three search \textit{strategies} and then reported the number of outliers it found as a fraction of its \textit{capability}. Then, the subject reported their total outlier estimate. Human subjects then evaluated statements about the agent's behavior, reliability, and their trust in the agent.

\subsubsection{Results}
\includedSubjects{} subjects were recruited. Self-reported trust was modeled using Ordinary Least Squares, but the group that interacted with varying \textit{capability} agents on a short time order produced a better performing ARIMAX model. Models were cross-validated between groups and found a moderate improvement in the next trial trust prediction.

\subsubsection{Conclusion}
A time series modeling approach reveals the effects of temporal ordering of agent performance on estimated trust. Recency bias may affect how subjects weigh the contribution of strategy or capability to trust. Understanding the connections between agent behavior, agent performance, and human trust is crucial to improving human-robot collaborative tasks.

\subsubsection{Application}
The modeling approach in this study demonstrates the need to represent autonomous agent characteristics over time to capture changes in human trust.

\subsection{Précis}
This study explores dynamic trust in a human-autonomy-teaming task, employing a screen-based activity with varying autonomous agent \textit{capabilities}. Findings indicate that temporal ordering of agent performance influences trust estimation. Time series models, particularly ARIMAX, enhance trust prediction, emphasizing the importance of considering temporal aspects in understanding human-robot collaboration.

%% file: sections/1introduction.tex
\begin{figure}[ht]
    \centering
    \includegraphics[width=0.5\textwidth]{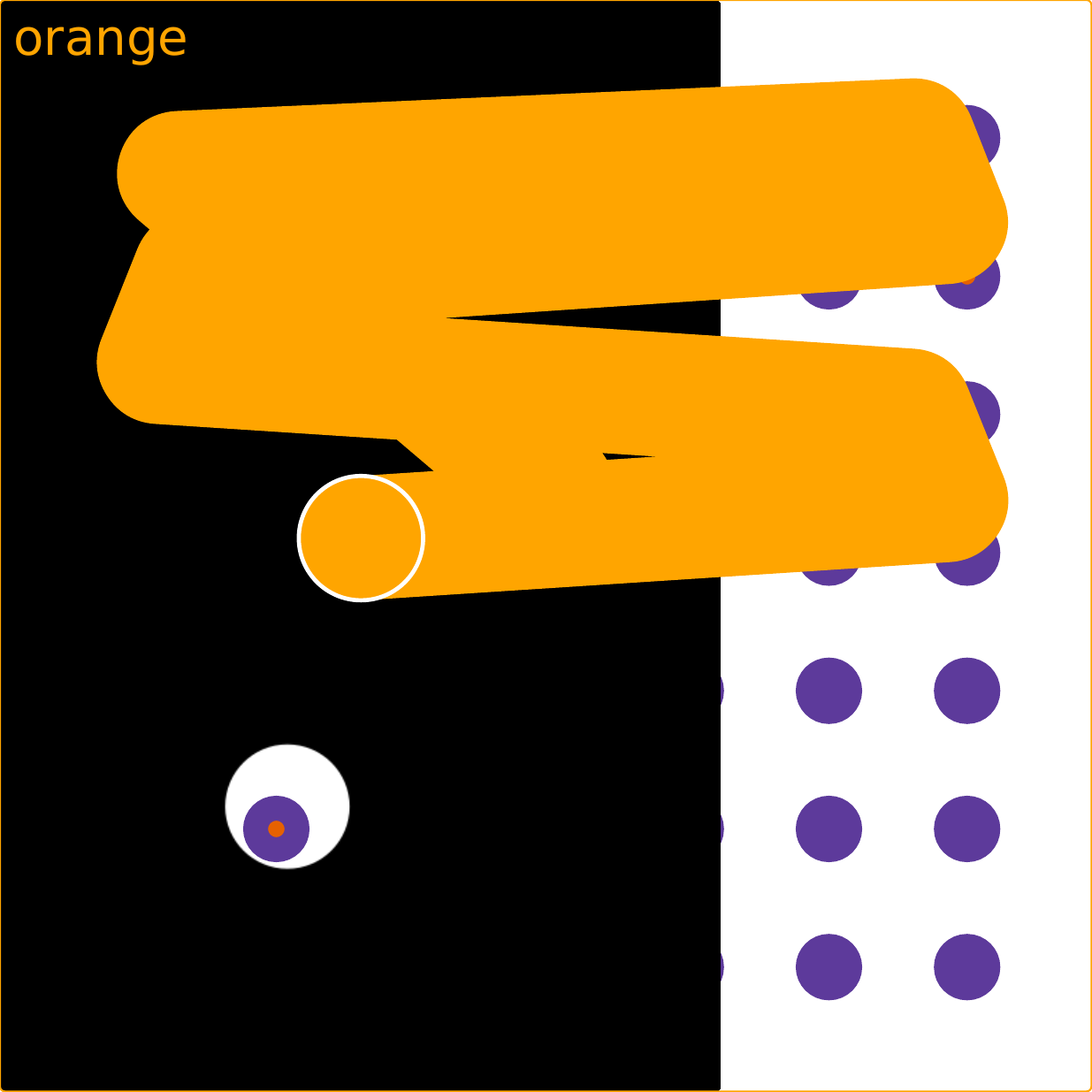}
    \caption{In the experimental task, the subject was instructed to search a hidden area for outlier circles: those with contrasting centers. The cutaway region on the right side shows a grid of possible outliers. Each trial, an Autonomous Searcher with differing \textit{capability} and \textit{strategy} searched alongside the subject. In this sample, the orange autonomous searcher used the "lawnmower" \textit{strategy}.}
    \label{maingame}
\end{figure}

Collaboration between humans and autonomous systems is changing every day. The rapid progress and integration between robotics and powerful deep learning technologies necessitates an understanding of how humans interact with agents more generally. Machines no longer only replace repetitive tasks for humans \cite{parasuraman_humans_1997}, but supplement and hasten decision-making. Humans must learn to work with autonomous systems to complete tasks with performance better than the sum of their parts. This study examines the time evolution of a human autonomous agent collaborative task and how changes in the order of agent performance change how human trust evolves.

Relying on an automated system to perform a task typically requires a human to supervise its performance. Depending on the performance of the automated system and environmental factors, the human's trust in the system changes \cite{lee_trust_1992}. Lee and See presented a definition of trust that is widespread in the literature: \textit{the attitude that an agent will help achieve an individual’s goals in a situation characterized by uncertainty and vulnerability} \cite{lee_trust_2004}. Trust is a latent variable and cannot be measured directly, but reliance actions are often used as a proxy \cite{desai_effects_2012,hu_computational_2019,okamura_adaptive_2020}. Although this definition encapsulates what it means to trust, more work has expanded on a multitude of components that can affect the attitude, like human predisposition \cite{hoff_trust_2015} or anthropomorphism \cite{cohen_anthropomorphism_2023}.

With an estimate of reliance and the underpinning that is trust, automated systems can adapt to the predicted behavior of a human. In this way, automated systems can change their level of autonomy \cite{parasuraman_model_2000}. When supervision isn't required, systems are considered autonomous and the paradigm of interaction changes \cite{endsley_toward_1995}. Autonomous systems can operate on longer interacting time scales that span hours or longer in which goal-defining or norm setting is negotiated between agents \cite{chiou_trusting_2023}. For these time scales, we must understand how trust in these relationships changes over time.

The experiment in this study pairs a human and autonomous agent in a 2d grid search task with changing autonomous agent \textit{strategy} and \textit{capability}. The human subject has the opportunity to observe the behavior of the autonomous agent while simultaneously searching the same environment (see Figure \ref{maingame}). Subjects answer a survey about their trust level and use information from the autonomous agent to estimate the number of outliers in the grid.

Subjects in the experiment are split into two groups, which, over the course of many trials, interact with agents whose \textit{strategy} and \textit{capability} parameters change on different timescales. Trust in autonomous agents depends on the performance of the agent in the task, but when subjects see varying performance (separated by embodiment), their analysis of the system as a whole may be required.

%% file: sections/2background.tex
\subsection{Trust in automation and autonomous systems}

Automatic systems are designed to control a process despite disturbances and manage setpoint changes. Those who use automated systems supervise the output's behavior to assess system failure. Depending on the person's expertise, they may rely on the system in counterintuitive ways. Parasuraman describes differences in reliance behavior in people with different expertise \cite{parasuraman_humans_1997}. Without a proper understanding of the automation system, people may \textit{misuse}, or overrely, or \textit{disuse}, neglect the system. When people can replace an automated system \cite{parasuraman_model_2000} or have the appropriate workload to supervise \cite{endsley_toward_1995}, actions they take that reduce the autonomy of the system may signal diminished trust in the system.

Lee and See suggest that measuring trusting behaviors does not necessarily indicate a trusting attitude \cite{lee_trust_2004}. Only when one analyzes the entire situation can a trusting action be attributed to an attitude of trust \cite{hoff_trust_2015,oleson_antecedents_2011}. When interactions become more complex, agent performance may \cite{desai_effects_2012} or may not \cite{robinette_effect_2017} lead to increases in trust. Understandably, measuring automation reliance is often tenable \cite{kohn_measurement_2021}, but without careful consideration, it may just measure reliance.

Autonomous systems are capable of working without human interference. Trust plays a different role since human interactions with autonomous systems can be at a higher decision-making level \cite{chiou_trusting_2023}. Measuring human trust while \textit{using} \cite{parasuraman_humans_1997} an automatic system must be translated to trust while \textit{working alongside} an autonomous system. To capture more than reliance-based trust, this paper uses self-reported trust \cite{jian_foundations_2000}, a subject's introspection, to understand how their characterization of an autonomous agent and the situation changes over time.

\subsection{Time series trust models}

Automated controllers do not have an anthropomorphized embodiment and no sense of personality. An automated controller is designed to do one task well, usually abiding by some physical or fundamental law. Supervisory control, where the human can enable automatic control of a system, can be influenced by an internal calculation that maximizes the task goal. If the system is better than the human, the human should choose the automatic system. The evaluation of system performance is a skill in itself. \cite{lee_trust_1992} collects trust survey responses while working with a semi-automatic pasteurization plant. By design, an automatic control system will perform within some bounds unless a sufficient disturbance causes unstable deviation from the setpoint. This disturbance may be a broad set of external factors the human collaborator can evaluate. For example, in this experiment, a pump failed for a trial on one day and all trials on another day. Without explanation, the human can only use performance to judge their trust or reliance on the system, which is a self-calibrated threshold. Subjects rated their trust in the system on a ten-point scale with specific instructions for what trust meant:``You will be asked to assess the plant's performance based on four criteria: the system's predictability, the system's dependability, the faith you have in the system and the amount of trust you have in the system." This may bias subjects to correlate their trust along the system's performance without regard for externalities.

The study used an ARMAV model of the form:
\begin{multline}
Trust(t) = \phi_{1}Trust(t-1)+A_{1}Performance(t) \\
+A_{1}\phi_{2}Performance(t-1) \\ +A_{2}Fault(t)+A_{2}\phi_{3}Fault(t-1)+a(t)
\end{multline}

Autonomous system performance can vary over several interactions with a human. The magnitude of trust change due to agent performance is a crucial measure for evaluating future interactions. Hu et al. used a driving simulator to measure trust levels after automatic obstacle detection failures \cite{hu_computational_2019}. Their state-space model used Expectation Bias, Cumulative Trust, and Experience to predict the next trust state. Given the population response, they measured the probability that a subject would choose to rely on automation. They found demographic specific parameterizations of a state-space trust estimator. In another car driving task, Desai et al. allowed subjects to change the level of autonomy in a remote control car \cite{desai_effects_2012}. They found subjects lowered the autonomy level of the car when reliability dropped, as well as a lower mean trust response. Similarly, Liu et al. used a driving task where the subject could indicate that they wanted to take over driving to indicate a decrease in trust \cite{liu_clustering_2021}. The environment was complicated by reduced visibility and diminished autonomous driving reliability. Subjects were fit into "skeptical" and "confident" clusters based on a portion of their interaction, which allowed the researchers to predict their trust using personalized models. 

Clustering-based trust studies are gaining popularity in trust time series with \cite{guo_modeling_2021} and \cite{bhat_clustering_2022} using Bayesian time series models. Depending on features of the subject's trust trajectories, subjects were clustered into "Bayesian decision maker," "oscillator," and "disbeliever" classes. The testbeds used simulated robotic decision support agents to identify risks in simulated buildings. Subjects rated their trust at their own discretion and after trials.

Although the dynamics of trust as a state have been studied in the mentioned references, with further examples in \cite{rodriguez_rodriguez_review_2023}, this study aims to uncover the effects of changing performance and behavior simultaneously at different time-scales.

%% file: sections/3experiment.tex
\subsection{Experimental Task}
The goal of the experimental task was for subjects to interact with an autonomous agent that exhibited varying characteristics. Then, subjects could report measures of the agent's performance and an introspection of their trust in the agent. We designed the ordering of agent characteristics as experimental conditions to account for multiple conditions simultaneously. The experiment was conducted online through a web browser, allowing the test of many subjects outside the lab.

We instructed subjects to search a hidden area simultaneously with an Autonomous Searcher (AS). They did so with a "spotlight," which provided a view through an opaque mask to a grid of potential "outliers" seen in Figure \ref{maingame}. Subjects were to report the total number of outliers in the grid. The AS was embodied as an opaque circle but implied the ability to view circles and outliers below the mask. It left a persistent trail to display the previously searched areas in the current trial. The grid was a seven-by-seven array of solid purple circles with outliers containing an orange center. In each trial, a different number of outliers between five and fifteen were randomly placed in the grid.

\begin{figure}[hb]
    \centering
    \includegraphics[width=\columnwidth]{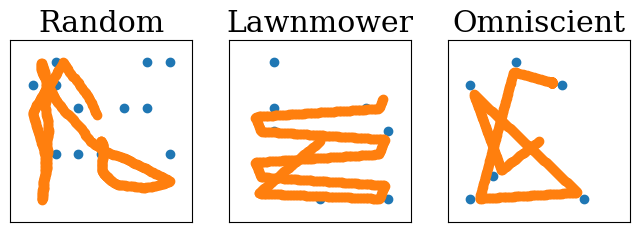}
    \caption{Sample AS paths from each \textit{strategy} shown in orange. The paths shown are eroded to illustrate the behavior better. Outliers are blue circles. The Random and Omniscient paths are generated randomly but shared within subjects of the same group.}
    \label{searcherpath}
\end{figure}

The \textit{capability} of the AS varied according to its color: blue, orange, and yellow. The grid area displayed the color name in case of subject colorblindness. The AS kept a running count of the outliers that it intersected and reported a fraction of the total at the end of the trial: 20\%, 50\%, or 100\%. The \textit{capability} of agents was not told to subjects but could be discovered by color. Additionally, AS's used one of three \textit{strategies} a given trial. The first, "Lawnmower," moves back and forth across the area in a predictable and identical pattern. Next, the "Random" \textit{strategy} moved with a uniformly distributed force input each time step. Lastly, the "Omniscient" \textit{strategy} moved directly to each outlier, which appeared similar to the "Random" \textit{strategy} but performed very well. See Figure \ref{searcherpath} for examples of search \textit{strategies} generated in the study.

The subject controlled their spotlight using the arrow keys on their keyboard. On each frame during the game, velocity accumulated in the same direction for each arrow key they held up to a maximum speed. The spotlight speed decreased if they did not enter a control input for an axis during a frame. When the spotlight contacted a wall, the velocity was set to 0 in that direction. This form of force control with damping was chosen due to the reported ease of learning during pilot testing. The spotlight and AS sizes were smaller than the distance between grid circles to prevent searching multiple rows or columns simultaneously.

Each trial expired after a 20-second time limit. This time is 80\% of the duration it would take to optimally encounter each grid circle, limiting the amount of information the subject could gain from searching alone. Subjects received a 5-second warning before the trial ended.

\begin{table*}[h]
    \input{figures/factor_table}
    \caption{Subjects are assigned to one of two groups, and each trial searches with an AS of a given \textit{capability} using a given \textit{strategy}. At the end of each trial, subjects receive a survey with the above measures. The relationship between measures was studied in \protect\cite{dekarske_human_2021}.}
    \label{anova_measures}
\end{table*}

\subsection{Inter-Trial Questionnaire} \label{questionnaire}

After the end of each trial, subjects received a two-part questionnaire. The first asked two questions to evaluate their performance in the task and the outcome of their outlier count and application of AS count to the total estimate. The second part asked the subjects to self-report their trust in the AS.

The first question, "How many outliers did you find with your spotlight this trial?" was intended to measure how well subjects could search and contribute to measuring experiment compliance. The interface then displayed a report from the AS: \textit{The \textbf{color} autonomous searcher reports finding \textbf{x} outliers}. Second, subjects were asked, "How may total outliers were hidden in the entire grid this trial?"

If the subject decided to split the search effort evenly with the AS, they could use the report of AS-found outliers and their own found outliers to reach an accurate estimate. However, without knowing the exact \textit{capability} of the AS, the estimate would be more or less valuable. This could lead subjects to overlap their search until they found the performance of the AS or increase their reported outlier estimate to compensate.

After completing the task survey, the page displayed the number of outliers in the search area and the subject's cumulative score. The score was a function of the outlier error used to help motivate subjects. Then, subjects received a trust survey,  adapted from \cite{jian_foundations_2000}:

\begin{itemize}
  \item Statement 1: I am familiar with the autonomous searcher's strategy. (Strategy-focused Measure)
  \item Statement 2: The autonomous searcher is reliable. (Capability-focused Measure)
  \item Statement 3: I trust the autonomous searcher. (Trust Measure)
\end{itemize}

Subjects scored each statement on a Likert scale from 1-"Not at All" to 9-"Extremely." The survey was reduced in question count from the original survey to reduce repetition throughout the experiment. Additionally, each statement corresponded to a varying factor in the experiment - \textit{strategy} and \textit{capability}.

\subsection{Experimental Design}

The experiment was written in Javascript using the D3.js library to render an SVG container in a web browser. The game loop and logging ran at 30 frames per second. All game data was served from and stored on a laboratory server. The University of California's Davis Institutional Review Board approved this research, and it complies with the American Psychological Association Code of Ethics. Participants provided consent and were allowed to exit the experiment without penalty.

The game guided Subjects through an interactive tutorial, which introduced the task elements one at a time—first, spotlight movement and the time limit before introducing an AS as a teammate. Subjects were reassured that an AS was helpful, could vary in its ability to detect outliers, and used different movement \textit{strategies}. Training took approximately 3 minutes. Then, subjects searched alone for nine trials to practice and measure their ability to search.

\begin{figure}[h]
    \centering
    \includegraphics[width=\columnwidth]{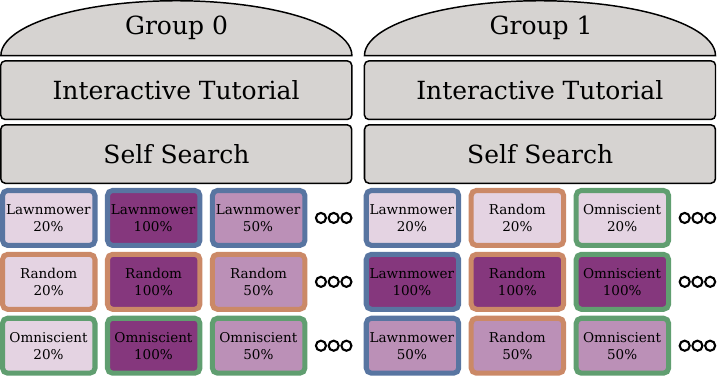}
    \caption{Subjects were assigned to one of two groups according to the order in which they began the experiment. Both groups started with an identical tutorial and self-search block before starting the main experiment. Subjects in group 0 had a constant \textit{strategy} in each block, while group 1 had a constant \textit{capability} in each block. In this figure, \textit{capability} is indicated by the shade of the box and \textit{strategy} by the outline of the box.}
    \label{design}
\end{figure}

The remaining 63 trials were sorted into three blocks of 21. The total number of trials was found from pilot testing to be reasonable before subjects became disinterested. Figure \ref{design} depicts a flow chart for how subjects encountered each factor depending on their assigned group. Group 0 subjects saw a constant \textit{strategy} among \textit{lawnmower}, \textit{random}, and \textit{omniscient} each block while group 1 saw a constant \textit{capability} 20\%, 50\%, or 100\%. Within each block, conditions of the remaining factor were varied. A random seed, shared with all subjects, ensured that AS motion, outlier count, outlier position, and trial order were the same across subjects of the same group. The study variables are listed in Table \ref{anova_measures}.

%% file: figures/factor_table.tex
\begin{tabular}{@{}rlp{0.6\textwidth}@{}}
\multicolumn{3}{l}{Factors}                                                                         \\ \midrule
Strategy                    & Lawnmower        & Moves across columns, one row at a time. Frequently used by subjects in pilot experiments. The most predictable strategy. \\
            & Random               & Moves randomly, obeying the dynamics that the spotlight uses.  \\
            & Omniscient           & Moves to each outlier in a random order.                       \\ \midrule
\multirow{3}{*}{Capability} & 20\%             & \multirow{3}{=}{The percentage of outliers found that are reported to the subject at the end of each trial.}              \\
            & 50\%                 &                                                                \\
            & 100\%                &                                                                \\ \midrule
\multirow{2}{*}{Group}      & Blocked Strategy & Each block has a constant strategy with varying capability.                                                               \\
            & Blocked Capability   & Each block has a constant capability with varying strategy.    \\ \midrule
\multicolumn{3}{l}{Measures}                                                                        \\ \midrule
Familiarity                 & \multicolumn{2}{l}{A measure designed to test the predictability of the AS.}                          \\
Reliability & \multicolumn{2}{l}{A measure which focused on the task performance of the AS.}        \\
Trust       & \multicolumn{2}{l}{The main focus of the study which encapsulates the prior metrics.} \\ \bottomrule
\end{tabular}

%% file: sections/4trustmodel.tex
\begin{figure}[h]
  \centering
  \input{figures/modeling_process}
  \caption{This study follows the modeling process outlined in this flowchart.}
  \label{modeling_process}
\end{figure}
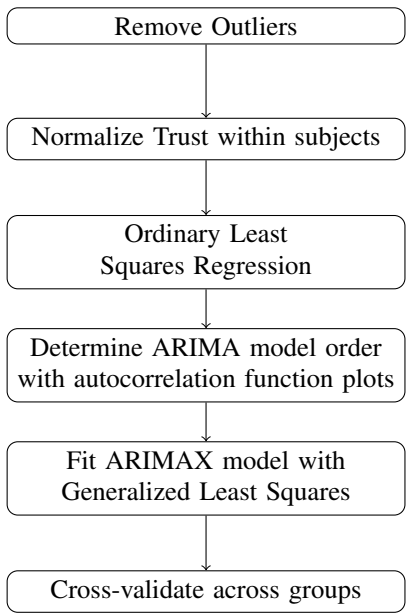

In a previous work, we performed an analysis of variance to find how survey factors relate to trust when considering each trial independently \cite{dekarske_human_2021}. We found that trust was correlated with \textit{capability}, and more legible \textit{strategies} resulted in higher trust survey ratings. The goal of this work is to demonstrate the influence of autonomous agent behavior and performance on a human interactor's trust in time. The modeling process follows the steps outlined in Figure \ref{modeling_process}.

Subjects' trust ratings varied in level and variance. To capture the extent of subjects' trust ratings compared to the population, their scores were normalized according to

\begin{equation}
    trust_{normalized} = \frac{trust - trust_{min}}{trust_{max}-trust_{min}}
\end{equation}

This normalization frames trust ratings concerning their best and worst experiences with an autonomous agent. The range could be compared to a subject's predisposition to trust.

If we assume that a subject's trust rating depends only on the \textit{capability} of a given agent, we could model trust with only knowledge of the \textit{capability} of an agent at that time, using a multiple categorical regression:
\begin{equation}
    y_t = \beta_0 + \beta_1 x_{1,t} + \dots + \beta_k x_{k,t} + \epsilon \\
    \label{regression}
\end{equation}
where $y_t$ is an estimate of reported trust at time $t$,  $\beta_k$ is the coefficient for the $k$th exogenous variable $x$, and $\epsilon$ is the error term. We assume that $\epsilon \sim N(0,\Sigma)$ with constant $\Sigma$ and not autocorrelated.

Lee and Moray showed a trust model with autoregressive components, suggesting that $\epsilon$ is likely autocorrelated \cite{lee_trust_1992}. A human's evaluation of the system, the agents, their environment, and the task changes throughout the experiment \cite{hoff_trust_2015}, which suggests that their trust evaluation changes as well.

Humans potentially carry separate trust state evaluations of each agent in any experiment \cite{williams_deconstructed_2021}. However, the homogeneity of the agents suggests shared traits that bias trust evaluations across agents. Thus, capturing agent-specific and system-general contributions to a model describing how trust changes over time is crucial.

We assume that since \textit{capability} is assigned to specific agents by color, they can be represented by their \textit{capability}. As subjects interact with an agent, they learn about its abilities and generate a level of \textit{perceptual trust} in it \cite{cohen_anthropomorphism_2023}. However, since the experiment presents each agent in the same way (between subject groups), the subject's change in trust introspection over time can be captured by a \textit{system-general} time process.

Assuming that a subject's learned trust changes similarly for each agent, we can expand the model in Section \ref{regression} to include time components in the error. Here, we use an ARIMA model with exogenous regressors of the form from \cite{hyndman_forecasting_2018}:
\begin{align}
    y_t &= \beta_0 + \beta_1 x_{1,t} + \dots + \beta_k x_{k,t} + \eta_t \\
    (1-\phi_1B)(1-B)\eta_t &= (1+\theta_1B)\varepsilon_t
\end{align}
where $\eta$ is the regression error, $B$ is the backshift operator ($B^{k}y_{t}=y_{t-k}$), $\phi$ is an autoregressive coefficient, $\theta$ is a moving average coefficient, and $\epsilon$ is the error term.

An ARIMA model consists of autoregressive, differencing, and moving average components. An autoregressive term measures the correlation with prior values of the predicted variable, meaning an estimate of trust depends on past measures of trust with some weight. Differencing removes non-stationarity from a time series. Differencing a trust time series corresponds to a correction on factors that influence trust in time. The moving average component measures the contribution of prior error terms to the predicted variable. In this case, the component can weight a trust estimate that differs from prior predictions.

To obtain the values of the parameters of the model, we build the ARIMA model and fit using loglikelihood maximization in the Python package Statsmodels \cite{seabold_statsmodels_2010}. We check the partial autocorrelation function plots and residual plots to check the efficacy of a model. Then, models are crossvalidated across groups using root-mean-square-error of the residuals from a one-step-ahead forecast of trust.

%% file: figures/modeling_process.tex
\begin{tikzpicture}[
    node distance = 1.5cm,
    every node/.style={rectangle, draw, rounded corners, text width=5cm, minimum width=1cm, align=center},
]

    \node (item0) {Remove Outliers};
    \node (item1) [below of=item0] {Normalize Trust within subjects};
    \node (item2) [below of=item1] {Ordinary Least Squares Regression};
    \node (item3) [below of=item2] {Determine ARIMA model order with autocorrelation function plots};
    \node (item4) [below of=item3] {Fit ARIMAX model with Generalized Least Squares};
    \node (item5) [below of=item4] {Cross-validate across groups};

    \draw[->] (item0) -- (item1);
    \draw[->] (item1) -- (item2);
    \draw[->] (item2) -- (item3);
    \draw[->] (item3) -- (item4);
    \draw[->] (item4) -- (item5);

\end{tikzpicture}

%% file: sections/5results.tex
\subsection{Demographics and exclusion}\label{demographics}

Subject data were excluded from the analysis if they did not complete the entire study, took out-of-protocol breaks, or answered task survey questions unreasonably. There were \includedSubjects{} total subjects included, of which \male{} identified as male, \female{} female, and \nonbinary{} nonbinary. The average age was \avgAge{} with standard deviation \stdAge{}. Of the \games{} subjects who played video games, the median subject played between \amountGames{}.

\subsection{Time series modeling}\label{timeseriesmodeling}

\begin{figure}[h]
  \centering
  \includegraphics[width=\linewidth]{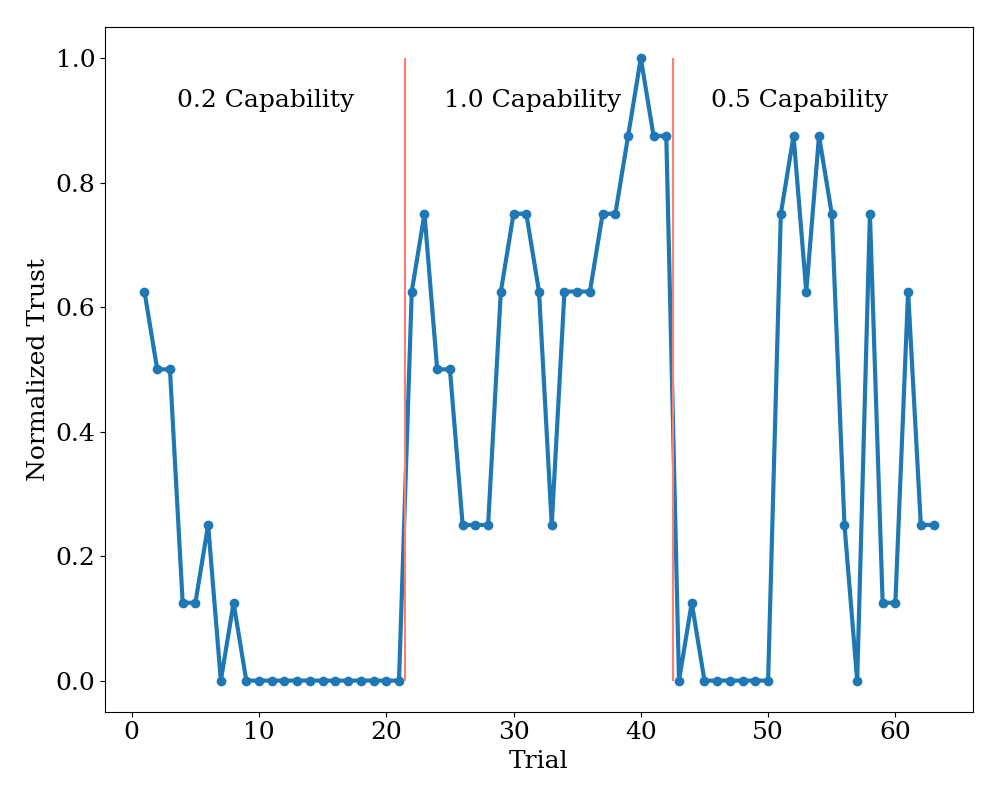}
  \caption{Single subject representative trust trajectory.}
  \label{individual_trust}
\end{figure}

Subjects completed 63 trials each, reporting their trust at the end of each trial. We can construct a discrete time series from these survey results that tracks survey responses throughout the experiment. Individual trust trajectories are sometimes erratic and vary to the extent of their normalized trust range. Figure \ref{individual_trust} shows a representative subject's ratings. As mentioned in Section \ref{search_behavior}, within-trial interactions between the subject and searcher hold a richness of information that cannot be entirely captured in this rating and may be the reason for drastic jumps in the trust survey.

\begin{figure*}[h]
  \begin{subfigure}{0.5\linewidth}
    \centering
    \includegraphics[width=\linewidth]{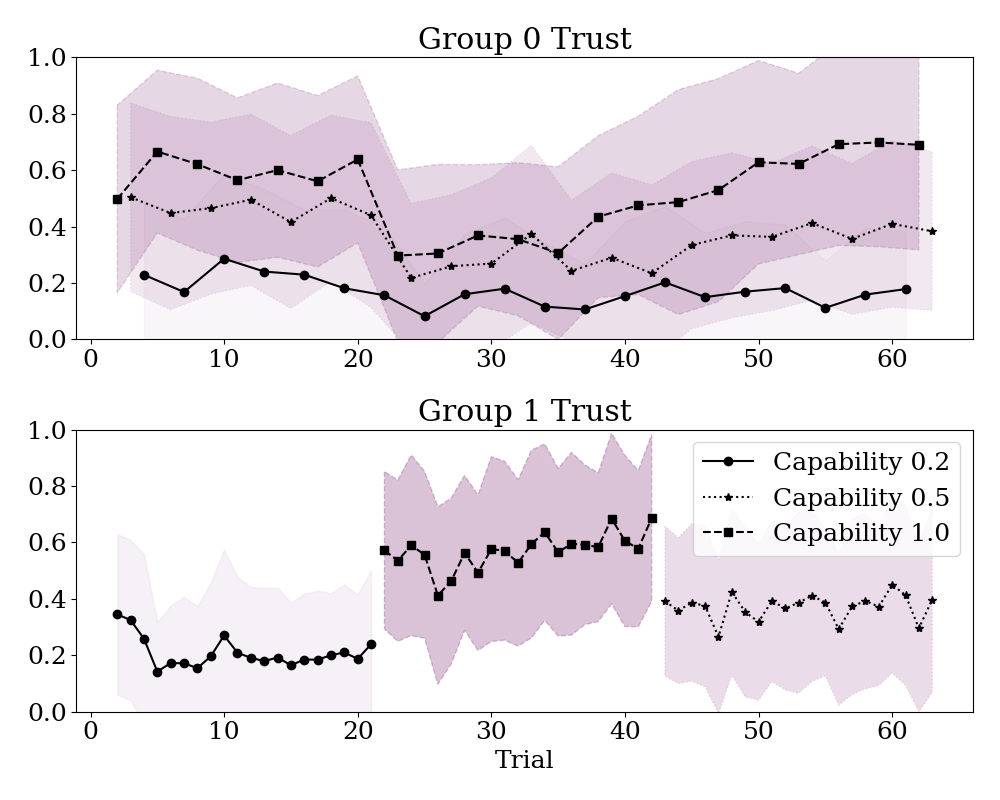}
  \end{subfigure}%
  \begin{subfigure}{0.5\linewidth}
    \centering
    \includegraphics[width=\linewidth]{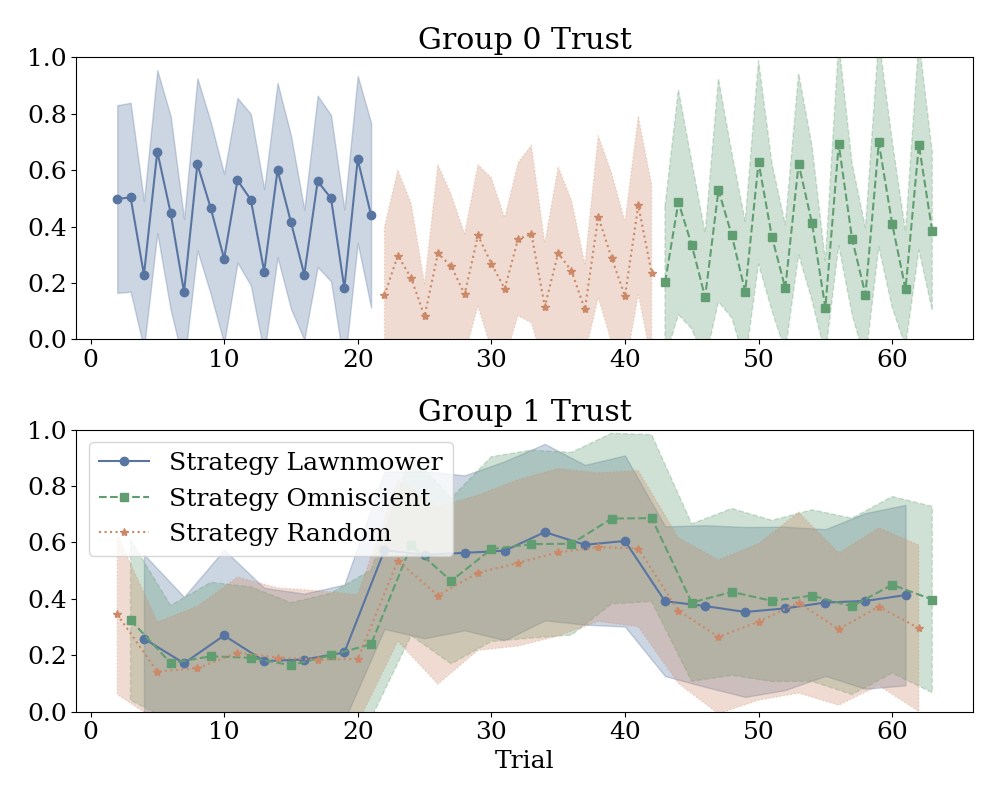}
  \end{subfigure}
  \caption{Each group's average trust ratings throughout the experiment. The data across rows are the same but colored and connected to emphasize the contributions of each experimental block to the rating. Since \textit{capability} likely contributes more to an individual's trust rating and Group 1 saw a constant searcher \textit{capability}, their ratings were more stable. The oscillations for Group 0 can be attributed to the varying performance of the searchers over a short frequency. The colors in this figure correspond to those in Figure \ref{design}.}
  \label{average_trust}
\end{figure*}

We are concerned with the effects of the autonomous searcher's \textit{capability} and \textit{strategy} on the subject's reported trust over time. Trust is, in part, tied to the task performance \cite{yang_toward_2023,desai_effects_2012,hu_computational_2019} of a human's partner. Part of the goal of the experimental design was to change the temporal focus of parameters. That is, vary the \textit{capability} of agents for group 0 every trial, but every block for group 1. Figure \ref{average_trust} compares each group's average trust trajectory split between the parameters. In the case of the \textit{capability} parameter, represented by different embodiments of the agents, one could interpret the trust ratings in the first column as the trust in each agent over time.

Given the differences in trust trajectories between groups, we construct separate models and cross-validate them to analyze the model's generality. Any differences in modeling may point to the importance of how parameters other than autonomous agent \textit{capability} affect subjects' trust in it.

\begin{table*}[h]
\centering
\input{generated_figures/ols0}
\caption{Group 0 multiple linear regression.}
\label{ols0}
\end{table*}

\begin{table*}[h]
\centering
\input{generated_figures/ols1}
\caption{Group 1 multiple linear regression}
\label{ols1}
\end{table*}

We start with a multiple linear regression with categorical variables \textit{capability} and \textit{strategy}. Due to the experimental design, \textit{strategy}, and \textit{capability} are found to be collinear. Trust can largely be predicted from \textit{capability} alone in this process, and with \textit{strategy} included, it leads to large coefficients and potential overfit of the model. Using only \textit{capability} provides a model where the coefficients are the average group trust values for each \textit{capability}. Tables \ref{ols0} and \ref{ols1} show the outputs of an Ordinary Lease Squares fit from Statsmodels \cite{seabold_statsmodels_2010}. The RMSE for Group 0 is \OLSRMSEGzero{} and \OLSRMSEGone{} for Group 1.

\begin{figure*}[h]
  \begin{subfigure}{0.5\linewidth}
    \centering
    \includegraphics[width=\linewidth]{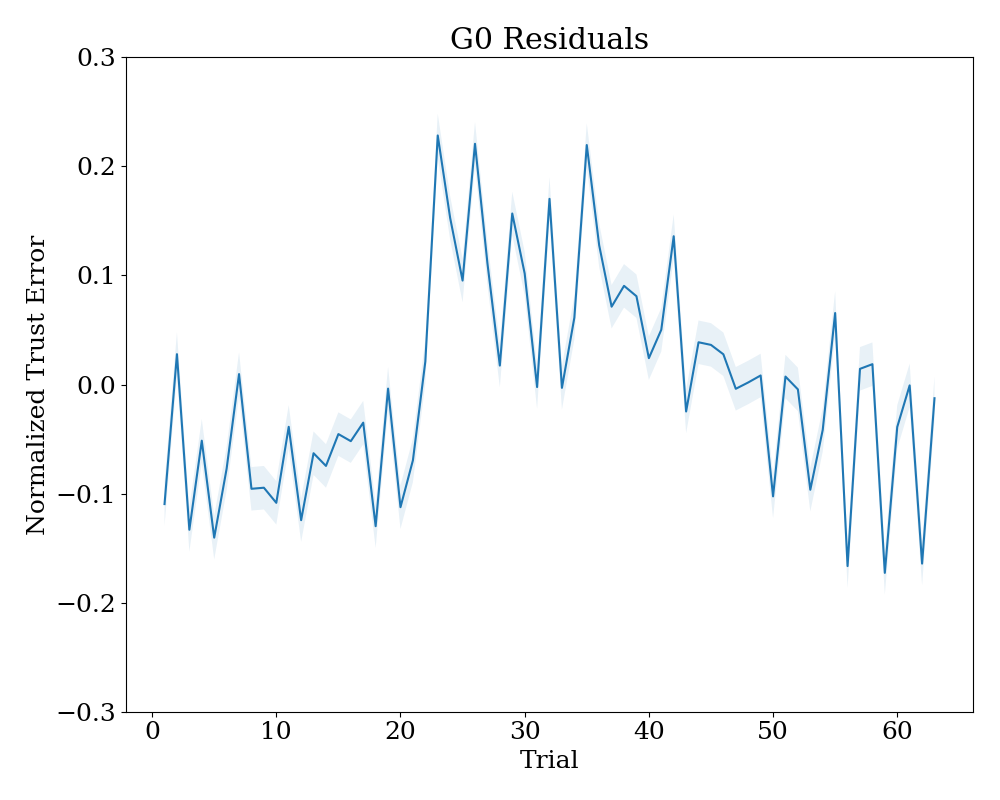}
  \end{subfigure}%
  \begin{subfigure}{0.5\linewidth}
    \centering
    \includegraphics[width=\linewidth]{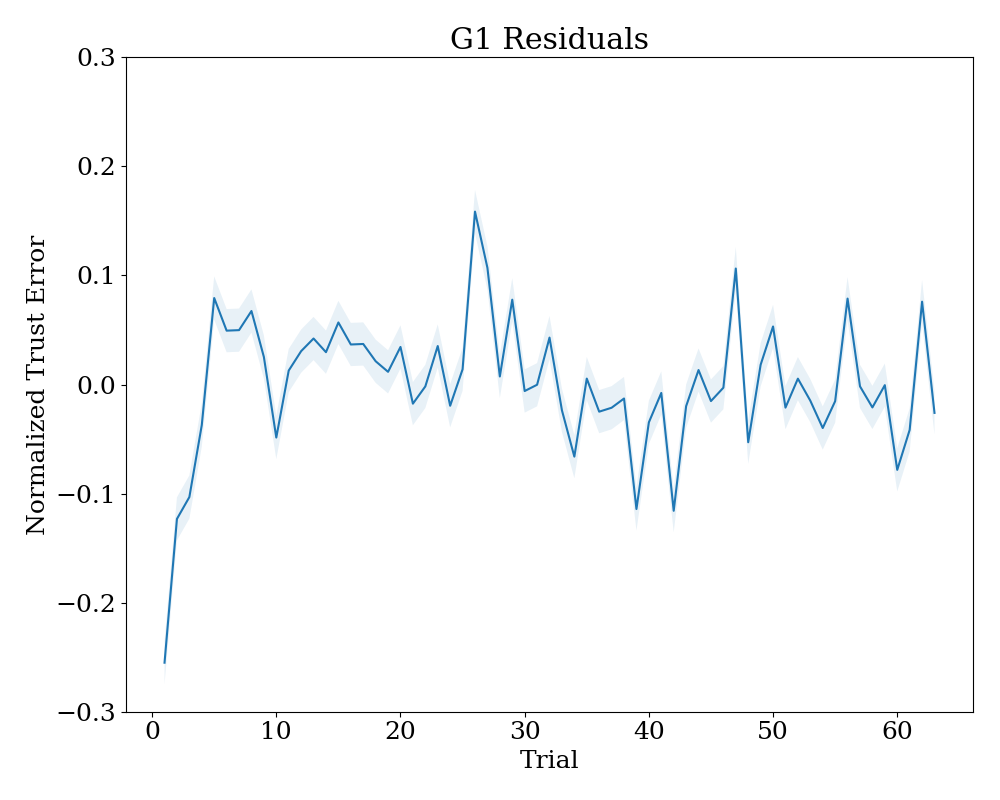}
  \end{subfigure}
  \caption{The multiple regression model residuals show strong non-stationarity and cycling in Group 0.}
  \label{ols_error}
\end{figure*}

\begin{figure}[h]
  \begin{subfigure}{0.5\linewidth}
    \centering
  \includegraphics[width=\linewidth]{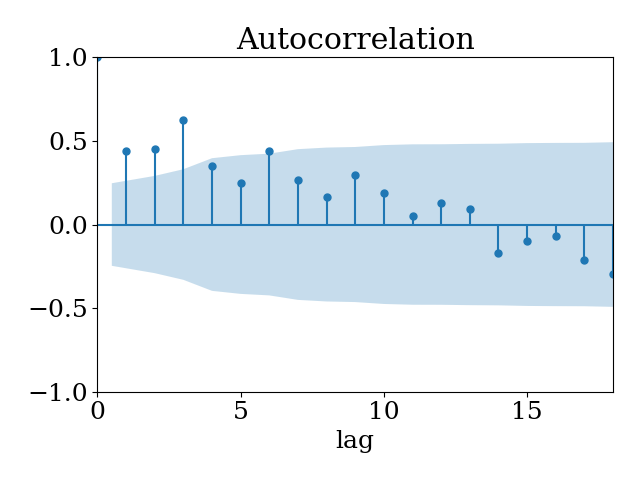}
  \end{subfigure}%
  \begin{subfigure}{0.5\linewidth}
    \centering
  \includegraphics[width=\linewidth]{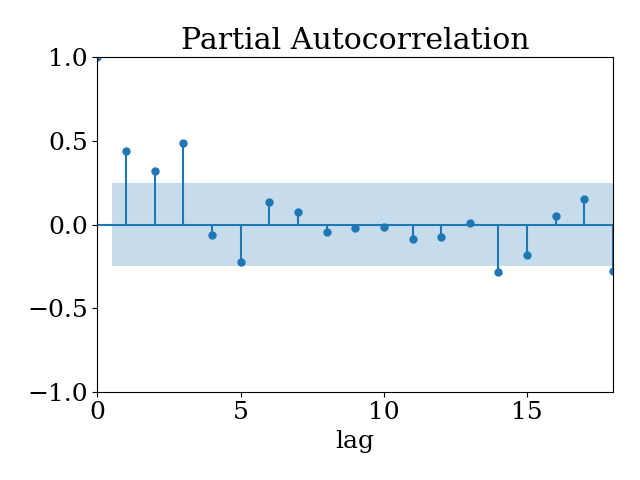}
  \end{subfigure}
  \caption{The Autocorrelation Function and Partial Autocorrelation Function plots of the residuals for Group 0 indicate lags 1 through 3 are likely necessary to capture the remaining error.}
  \label{acf_plot}
\end{figure}

\begin{table}[h]
\centering
\input{generated_figures/aic0.tex}
\caption{The array of AIC values calculated for different ARIMA(p,1,q) model orders over group 0 data. \textbf{p} values are the rows and \textbf{q} values are the columns. The results indicate that an ARIMA(2,1,3) is the most suitable model order given the goodness of fit and number of parameters.}
\label{AIC}
\end{table}

The residuals indicate that for Group 0, a model that includes temporal components is necessary. To capture non-stationarity, we add a differencing component. Also, we perform a partial autocorrelation of the error to find the order of the autoregressive component. Since lags 1, 2, and 3 have magnitudes above the critical value thresholds, at least a 3rd order model should be used. Iterating over combinations of Autoregressive ($p$) and Moving Average ($q$) orders with 1st order differences results in the lowest Akaike Information Criterion (AIC) shown in table \ref{AIC}.

\begin{table*}[h]
\centering
\input{generated_figures/arimax0.tex}
\caption{Coefficient table for group 0. Ascending \textit{capability} coefficients agree with the assumption that trust increases with increasing \textit{capability}. Significant autoregressive coefficients suggest the time series has strong cyclic signals from the study design.}
\label{arimaxtable0}
\end{table*}

\begin{table*}[h]
\centering
\input{generated_figures/arimax1.tex}
\caption{Coefficient table for group 1. Ascending \textit{capability} coefficients agree with the assumption that trust increases with increasing \textit{capability}. Relatively large ARIMA coefficients suggest that this model may be overfitting.}
\label{arimaxtable1}
\end{table*}

\begin{figure*}[h]
  \begin{subfigure}{0.5\linewidth}
    \centering
    \includegraphics[width=\linewidth]{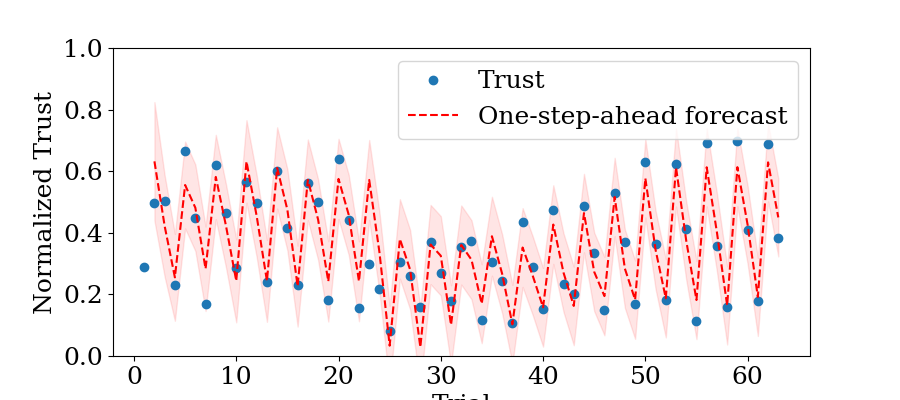}
  \end{subfigure}%
  \begin{subfigure}{0.5\linewidth}
    \centering
    \includegraphics[width=\linewidth]{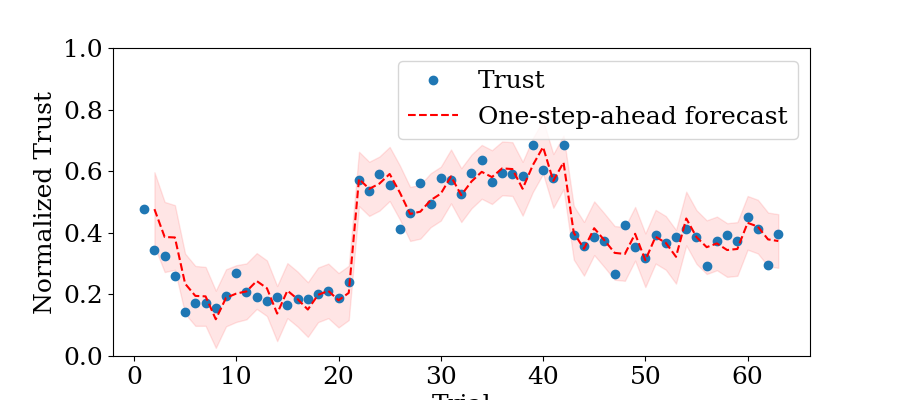}
  \end{subfigure}
  \caption{The one-step-ahead predictions for the ARIMAX model for group 0 (left) and group 1 (right).}
  \label{arimax}
\end{figure*}

\begin{table*}[h]
\centering
\begin{tabular}{rll}
\textbf{Group} & \textbf{0} & \textbf{1} \\
\textbf{Blocked} & \textbf{Strategy} & \textbf{Capability} \\
\midrule
\textbf{Linear Regression} & \OLSRMSEGzero{} & \OLSRMSEGone{} \\
\textbf{Group 0 ARIMAX} & \ARIMAXRMSEGzero{} & \ARIMAXRMSEGCROSSone{}\dag \\
\textbf{Group 1 ARIMAX} & \ARIMAXRMSEGCROSSzero{}\dag & \ARIMAXRMSEGone{} \\
\end{tabular}
\caption{Predicted vs. actual RMSE of time series models. \dag indicates cross-validated results.}
\label{RMSE}
\end{table*}

The final models for groups 0 and 1 describe the trends in the data as seen in Figure \ref{arimax}. The difference in \textit{capability} presentation in the experiment elicited the need for more complex modeling beyond linear regression. Including autoregressive components allows the model to capture information from prior trials. Cross-validating the results (table \ref{RMSE}) indicates an improvement in the linear regression model for both training models but a decrease in performance on the group 0 model cross-validated on group 1 data.

%% file: generated_figures/ols0.tex
\begin{center}
\begin{tabular}{lclc}
\toprule
\textbf{Dep. Variable:}    &     trust     & \textbf{  R-squared:         } &     0.182   \\
\textbf{Model:}            &      OLS      & \textbf{  Adj. R-squared:    } &     0.181   \\
\textbf{No. Observations:} &       2646    & \textbf{  F-statistic:       } &     294.0   \\
\textbf{Covariance Type:}  &   nonrobust   & \textbf{  Prob (F-statistic):} & 5.21e-116   \\
\bottomrule
\end{tabular}
\begin{tabular}{lcccccc}
                         & \textbf{coef} & \textbf{std err} & \textbf{t} & \textbf{P$> |$t$|$} & \textbf{[0.025} & \textbf{0.975]}  \\
\midrule
\textbf{capability\_0.2} &       0.1778  &        0.010     &    17.491  &         0.000        &        0.158    &        0.198     \\
\textbf{capability\_0.5} &       0.3710  &        0.010     &    36.490  &         0.000        &        0.351    &        0.391     \\
\textbf{capability\_1.0} &       0.5257  &        0.010     &    51.714  &         0.000        &        0.506    &        0.546     \\
\bottomrule
\end{tabular}
\end{center}

%% file: generated_figures/ols1.tex
\begin{center}
\begin{tabular}{lclc}
\toprule
\textbf{Dep. Variable:}    &     trust     & \textbf{  R-squared:         } &     0.201   \\
\textbf{Model:}            &      OLS      & \textbf{  Adj. R-squared:    } &     0.200   \\
\textbf{No. Observations:} &       2394    & \textbf{  F-statistic:       } &     299.8   \\
\textbf{Covariance Type:}  &   nonrobust   & \textbf{  Prob (F-statistic):} & 6.49e-117   \\
\bottomrule
\end{tabular}
\begin{tabular}{lcccccc}
                         & \textbf{coef} & \textbf{std err} & \textbf{t} & \textbf{P$> |$t$|$} & \textbf{[0.025} & \textbf{0.975]}  \\
\midrule
\textbf{capability\_0.2} &       0.2217  &        0.010     &    21.941  &         0.000        &        0.202    &        0.241     \\
\textbf{capability\_0.5} &       0.3716  &        0.010     &    36.785  &         0.000        &        0.352    &        0.391     \\
\textbf{capability\_1.0} &       0.5704  &        0.010     &    56.460  &         0.000        &        0.551    &        0.590     \\
\bottomrule
\end{tabular}
\end{center}

%% file: generated_figures/aic0.tex
\begin{tabular}{llrrrrrr}
\toprule
 &  & \multicolumn{6}{c}{q} \\
 &  & 0 & 1 & 2 & 3 & 4 & 5 \\
\midrule
\multirow[c]{6}{*}{p} & 0 & -99 & -128 & -129 & -127 & -131 & -132 \\
 & 1 & -115 & -127 & -127 & -127 & -130 & -128 \\
 & 2 & -138 & -136 & -136 & \cellcolor[HTML]{c0c0c0} \bfseries -142 & -140 & -138 \\
 & 3 & -136 & -134 & -136 & -140 & -138 & -136 \\
 & 4 & -135 & -134 & -133 & -137 & -132 & -134 \\
 & 5 & -136 & -134 & -133 & -135 & -133 & -132 \\
\bottomrule
\end{tabular}

%% file: generated_figures/arimax0.tex
\begin{tabular}{lcccccc}
                         & \textbf{coef} & \textbf{std err} & \textbf{z} & \textbf{P$> |$z$|$} & \textbf{[0.025} & \textbf{0.975]}  \\
\midrule
\textbf{capability\_0.2} &      -0.2060  &        0.020     &   -10.110  &         0.000        &       -0.246    &       -0.166     \\
\textbf{capability\_0.5} &      -0.0068  &        0.020     &    -0.344  &         0.731        &       -0.046    &        0.032     \\
\textbf{capability\_1.0} &       0.1385  &        0.019     &     7.477  &         0.000        &        0.102    &        0.175     \\
\textbf{ar.L1}           &      -0.7634  &        0.057     &   -13.356  &         0.000        &       -0.875    &       -0.651     \\
\textbf{ar.L2}           &      -0.9550  &        0.060     &   -15.806  &         0.000        &       -1.073    &       -0.837     \\
\textbf{ma.L1}           &       0.0437  &        0.181     &     0.241  &         0.809        &       -0.311    &        0.399     \\
\textbf{ma.L2}           &       0.6702  &        0.462     &     1.452  &         0.147        &       -0.234    &        1.575     \\
\textbf{ma.L3}           &      -0.5345  &        0.363     &    -1.473  &         0.141        &       -1.246    &        0.177     \\
\textbf{sigma2}          &       0.0041  &        0.002     &     1.817  &         0.069        &       -0.000    &        0.009     \\
\bottomrule
\end{tabular}

%% file: generated_figures/arimax1.tex
\begin{tabular}{lcccccc}
                         & \textbf{coef} & \textbf{std err} & \textbf{z} & \textbf{P$> |$z$|$} & \textbf{[0.025} & \textbf{0.975]}  \\
\midrule
\textbf{capability\_0.2} &      -0.1925  &        0.176     &    -1.091  &         0.275        &       -0.538    &        0.153     \\
\textbf{capability\_0.5} &      -0.1282  &        0.164     &    -0.784  &         0.433        &       -0.449    &        0.192     \\
\textbf{capability\_1.0} &       0.1395  &        0.096     &     1.453  &         0.146        &       -0.049    &        0.328     \\
\textbf{ar.L1}           &      -1.0205  &        0.056     &   -18.182  &         0.000        &       -1.131    &       -0.911     \\
\textbf{ar.L2}           &      -0.9945  &        0.023     &   -43.356  &         0.000        &       -1.039    &       -0.949     \\
\textbf{ma.L1}           &       0.8628  &        0.181     &     4.760  &         0.000        &        0.508    &        1.218     \\
\textbf{ma.L2}           &       0.7836  &        0.165     &     4.762  &         0.000        &        0.461    &        1.106     \\
\textbf{ma.L3}           &      -0.1097  &        0.159     &    -0.691  &         0.490        &       -0.421    &        0.201     \\
\textbf{sigma2}          &       0.0020  &        0.000     &     4.917  &         0.000        &        0.001    &        0.003     \\
\bottomrule
\end{tabular}

%% file: sections/6discussion.tex
\subsection{Subject search behavior depends on Autonomous Searcher conditions}\label{search_behavior}
\begin{figure}[h]
    \centering
    \includegraphics[width=\linewidth]{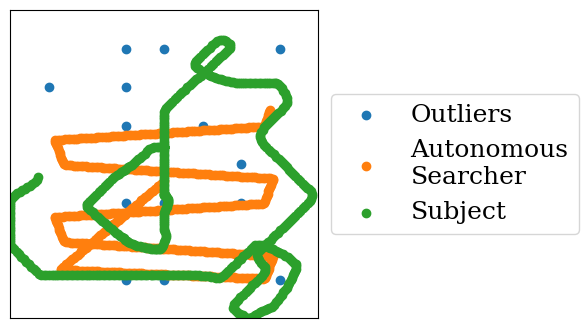}
    \caption{An example of the search trajectories of the human subject and autonomous searcher. They started in the center of the grid and searched similar areas before splitting. The subject overlapped the previously searched region near the bottom toward the end of the trial. The paths shown here are thinner than the spotlight's size during the task to better differentiate the searcher and subject.}
    \label{trial_example}
  \end{figure}%
  
\begin{figure}[h]
    \centering
    \includegraphics[width=\linewidth]{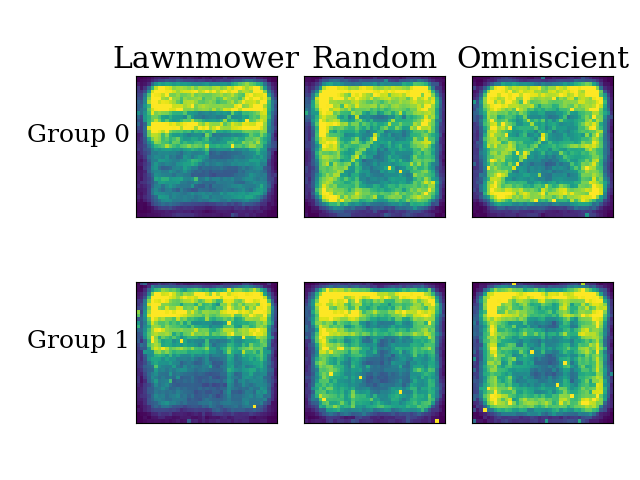}
    \caption{A heatmap of the subject searched area with the initial position removed. The Lawnmower \textit{strategy} was more predictable and resulted in subjects avoiding the autonomous searcher's area. The Random and Omniscient \textit{strategies} were more likely to search the center areas, resulting in subjects searching the exterior of the grid more often.}
    \label{heatmap}
\end{figure}

Subjects had 20 seconds to search the grid area with the autonomous searcher. Subjects displayed varied behaviors to collaborate with the autonomous searcher, as seen in Figure \ref{trial_example}. Some may have intentionally searched the same area to confirm the searcher's performance at the expense of their own search performance. Subjects from different groups may have learned the paths from the autonomous searcher over different time scales. Group 0 saw each \textit{strategy} repeated over the block, while Group 1 saw each \textit{strategy} rapidly. Depending on the subjects' propensity to consider the Autonomous Searchers \textit{strategy} or \textit{capability} (identity), the ordering of parameters may have affected their search behavior, as seen in Figure \ref{heatmap}. The analysis of subject search behavior is left for future work.

\subsection{Ordering effects captured using ARIMAX model}\label{ARIMAXDiscussion}

This study demonstrated how changing the frequency with which humans interact with differing \textit{capability} autonomous agents affects their trust. The subjects' trust changes from trial to trial according to the agent's \textit{capability} but cannot be predicted without knowledge of prior interactions. The dynamics of trust depend on aspects of the agent or the environment that change in time \cite{hoff_trust_2015}.

By having different blocked characteristics for each group, this experiment captured trust changes due to characteristics at a given time and the interaction between the changing characteristics over the experiment. Figure  \ref{average_trust}, in particular, emphasizes how reframing measured trust shows apparent temporal factors at play for a group's trust over the experiment. If trust was solely based on performance, we would see constant ratings when visualized as \textit{capability}. Group 0 saw three agents in rapid succession and could evaluate relative performance early in the experiment. However, since group 1 saw the same agent use different \textit{strategies}, they could quickly determine how the agent performed during the grid search task. The ARIMAX models developed here are a good candidate for these conditions since they account for the trust in time and the effect of agent properties.

An individual's trust at a given time is a function of their experience with the autonomous agent during prior interactions. Thus, the ARIMAX model, which describes contributions to trust estimation, suitably includes the exogenous factors that track the current situation and autoregressive factors from previous situations. The performance of the ARIMAX over the OLS models shown in Table \ref{arimax} agrees with the intuition that including autoregressive factors is necessary for capturing human trust evolution in time. The model selection process using AIC ensures that increasing the number of model parameters is appropriate, while cross-validation between groups suggests the model generalizes across experimental conditions. 

Inter-trial trust ratings capture a subject's trust in an autonomous agent throughout the experiment. It must be noted that when using trust surveys, subjects are biased to evaluate autonomous agents more critically. They may question, "If the experiment asks me if I trust the autonomous searcher, should I?". In this case, including separate autonomous agents means that their trust rating is grounded in evaluating the performance between agents. At the very least, the results are internally valid since the same trust survey was given to all subjects for all trials. Subjects evaluate autonomous agent performance during the trial by its behavior and after the trial by its outlier reporting. Since subjects were asked to think about their trust at this time, it can't be clear that their trust changed during the trial, if it did at all. Since autonomous agents report the number of outliers they found, there isn't a clear indication that they succeeded or failed at the task, matching the scale of trust presented in the survey. Task faults \cite{lee_trust_1992} are not always obvious in human autonomy teaming situations due to higher level interactions like goal negotiation \cite{chiou_trusting_2023}. 

Subjects could explore the region of the hidden region that the autonomous searcher already searched. After some trials, the subject could predict where the searcher could move due to the kinematics of the spotlights. However, the autonomous searcher did not adaptively search with the subject, which could improve task performance \cite{chen_planning_2018}. Future work includes allowing the autonomous agent knowledge of the human's trust, and giving the AS the ability to vary its search strategy to maximize task performance. If autonomous agents had additional actions to communicate intent \cite{dragan_legibility_2013} or \textit{capability} \cite{okamura_adaptive_2020}, trust would be aligned with task performance and could increase the rate at which a subject learned the task.

\subsection{Future trustworthy autonomous systems}
As autonomous systems gain capabilities, they are more likely to interact with humans with a similar decision-making level \cite{chiou_trusting_2023}. Considering how human trust changes in time could allow such systems to understand potential future human actions and plan accordingly. Although the presented questionnaire is intrusive, the system could ask if a system's confidence in the human trust state is low. Although not time optimal in the short term, taking time to better understand the situation could be vital in the long term \cite{cserna_planning_2017}. Findings from this experiment show promise in the ability to track human trust over time and the connection between autonomous agent characteristics and trust.

%% file: sections/7conclusion.tex
This study demonstrated how changing the frequency with which humans interact with differing \textit{capability} autonomous agents affects their trust. The subjects' trust changes from trial to trial according to the agent's \textit{capability}. However, a better prediction of their trust can be made with knowledge of prior interactions. The dynamics of trust depend on aspects of the agent or the environment that change in time \cite{hoff_trust_2015}.

Self-reported trust represents an introspection that subjects can make based on their evaluation of the current interaction environment. Future work can use this kind of self-reported measurement to estimate potential actions a human may take given their history of previous actions and trust ratings. Connecting human judgment of autonomous system capability and behavior with human action is crucial for moving the field of human autonomy teaming forward.

Humans interact with autonomous systems that are constantly improving in capability. Imbuing systems with empathy has the potential to make interactions more productive and comfortable.